\documentclass[10pt,twocolumn,twoside]{ieeetran}

\IEEEoverridecommandlockouts

\usepackage{graphics} 

\usepackage{epstopdf}
\usepackage{epsfig}
\usepackage{tikz}
\usepackage{color}
\usepackage{subcaption}
\usepackage{psfrag}
\usepackage{caption}
\usepackage{lipsum}
\usepackage{url}
\usepackage{amsmath,amsfonts, amssymb,color}

\newcommand{\Rb}{\mathbb{R}}
\newcommand{\xv}{\mathbf{x}}
\newcommand{\yv}{\mathbf{y}}

\renewcommand{\AA}{\mathcal{A}}

\newcommand{\GG}{\mathcal{G}}

\newcommand{\DD}{\mathcal{D}}

\newcommand{\UT}{\mathtt{U}}
\newcommand{\VT}{\mathtt{V}}

\allowdisplaybreaks

\usepackage[numbers]{natbib}
\usepackage{tikz}
\usetikzlibrary{positioning}
\usetikzlibrary{arrows}

\newtheorem{theorem}{Theorem}

\newtheorem{lemma}{Lemma}
\newtheorem{remark}{Remark}
\newtheorem{definition}{Definition}
\newtheorem{example}{Example}
\newtheorem{proposition}{Proposition}
\newtheorem{assumption}{Assumption}

\makeatletter
\renewcommand\bibsection%
{
  \section*{\refname
    \@mkboth{\MakeUppercase{\refname}}{\MakeUppercase{\refname}}}
}
\makeatother

\newcommand{\ignore}[1]{}

\newcommand{\oprocendsymbol}{\hbox{$\bullet$}}
\newcommand{\oprocend}{\relax\ifmmode\else\unskip\hfill\fi\oprocendsymbol}

\DeclareMathOperator*{\argmin}{argmin}

\title{Game-Theoretic Vaccination Against Networked SIS Epidemics and Impacts of Human Decision-Making}
\author{~Ashish~R.~Hota~and~Shreyas~Sundaram
\thanks{Ashish R. Hota is with the Department of Electrical Engineering, Indian Institute of Technology (IIT), Kharagpur, India. Shreyas Sundaram is with the School of Electrical and Computer Engineering, Purdue University, USA. E-mail: ahota@ee.iitkgp.ac.in, sundara2@purdue.edu. Part of this research was carried out when Ashish R. Hota was with the Automatic Control Laboratory, ETH Z{\"u}rich, Switzerland. This research was also supported in part by the National Science Foundation under grant no. CNS-1718637.}%
}

\begin{document}

\maketitle
\thispagestyle{empty}
\pagestyle{empty}

\begin{abstract}
We study decentralized protection strategies against Susceptible-Infected-Susceptible (SIS) epidemics on networks. We consider a population game framework where nodes choose whether or not to vaccinate themselves, and the epidemic risk is defined as the infection probability at the endemic state of the epidemic under a degree-based mean-field approximation. Motivated by studies in behavioral economics showing that humans perceive probabilities and risks in a nonlinear fashion, we specifically examine the impacts of such misperceptions on the Nash equilibrium protection strategies. We first establish the existence and uniqueness of a threshold equilibrium where nodes with degrees larger than a certain threshold vaccinate. When the vaccination cost is sufficiently high, we show that behavioral biases cause fewer players to vaccinate, and vice versa. We quantify this effect for a class of networks with power-law degree distributions by proving tight bounds on the ratio of equilibrium thresholds under behavioral and true perceptions of probabilities. We further characterize the socially optimal vaccination policy and investigate the inefficiency of Nash equilibrium.
\end{abstract}

\section{Introduction}
\label{section:introduction}
Cyber, physical, and social systems are becoming increasingly interdependent and interconnected. The security, robustness, and resilience of these large-scale networked systems depend on many factors, including the characteristics of attacks \citep{alpcan2010network,pastor2015epidemic}, topology of the network \citep{hota2016interdependent,drakopoulos2016network}, and centralized vs. decentralized decision-making \citep{alpcan2010network,hota2017impacts}. In addition, decisions made by humans that use these systems also have a significant impact on their security and resilience \citep{hota2017impacts,vanderhaegen2017towards,sanjab2017prospect}.

In this paper, we investigate the impacts of game-theoretic and human decision-making in the context of Susceptible-Infected-Susceptible (SIS) epidemics on networks \cite{pastor2015epidemic}. SIS epidemics have been shown to capture a wide range of dynamics, such as the spread of diseases in human society \citep{hethcote2000mathematics}, viruses in computer networks \citep{sellke2008modeling}, and opinions in complex networks \citep{lopez2008diffusion}. There is indeed a large literature on network epidemics, including mean-field approximations \citep{pastor2001epidemic,van2009virus}, characterizations of steady-state behavior \citep{khanafer2016stability,paarporn2017networked}, centralized protection strategies to control the spreading processes \citep{preciado2014optimal,drakopoulos2016network}, and network designs that are resilient against the epidemic \citep{trajanovski2017designing,hota2016optimal}; see \citep{nowzari2016analysis,pastor2015epidemic} for recent reviews. While centralized protection strategies are not practical for large-scale networked systems, decentralized and game-theoretic protection strategies against network epidemics have been relatively less explored \citep{nowzari2016analysis,pastor2015epidemic}. The existing literature has focused on two types of protection strategies; i) nodes choose their curing rates independently \cite{omic2009protecting,hota2017impacts}, and ii) nodes choose whether or not to vaccinate themselves \cite{theodorakopoulos2013selfish,bauch2004vaccination,trajanovski2015decentralized}. 

In this paper, we focus on the second class of protection strategies. The vaccination game has been studied in both non-networked settings (i.e., with a fully mixed population) \cite{theodorakopoulos2013selfish,bauch2004vaccination} and networked settings \cite{trajanovski2015decentralized}. In \cite{trajanovski2015decentralized}, nodes decide whether to purchase vaccination or risk being infected with probability given by the endemic state of the N-Intertwined Mean Field Approximation (NIMFA) \cite{van2009virus} of the SIS dynamics. The authors analyzed Nash equilibria in certain classes of networks, specifically, complete graphs, complete bipartite graphs, and multi-community networks. In a follow up work \cite{trajanovski2017designing}, the authors studied game-theoretic design of networks that are resistant against the SIS epidemic. 

In addition to being limited to certain classes of networks, a common assumption in the existing game-theoretic literature on epidemics is that the decision-makers are risk neutral (i.e., expected cost minimizers), and perceive infection probabilities as their true values. However, there is a large body of work in psychology and behavioral economics that has shown that humans perceive probabilities differently from their true values \citep{kahneman1979prospect,dhami2016foundations,barberis2013thirty} (see Section \ref{section:probweighting} for further details). Recent work has shown that these behavioral aspects of decision-making, specifically those captured by prospect theory \citep{kahneman1979prospect}, can have a significant impact on the efficiency, security and robustness of networked engineered systems \citep{hota2017impacts,hota2016fragility}. For instance, a recent stream of papers investigate various implications of prospect-theoretic preferences in the context of energy consumption decisions in the smart grid \citep{saad2016toward,etesami2018stochastic,xiao2015prospect}, and pricing in communication networks \citep{li2014users,yang2015prospect}.

Recent literature has also investigated the impacts of behavioral biases, such as mis-perceptions of probabilities, on the decisions made in the context of (network) security. In \citep{xiao2017cloud,xiao2018attacker}, the authors consider a cloud computing environment with a defender and a stealthy attacker, and investigate the pure and mixed Nash equilibria when both entities have (cumulative) prospect-theoretic preferences. They identify conditions under which behavioral biases of the attacker improve the security of the system. Impacts of nonlinear probability weighting were also studied for interdependent network security games \citep{hota2016interdependent} and network interdiction games \cite{sanjab2017prospect}.

Epidemic models are fundamentally different compared to the above settings. In contrast with \citep{xiao2017cloud,xiao2018attacker}, epidemics have multiple decision-makers captured by nodes in the network, and in contrast with \cite{hota2016interdependent}, epidemics spread through the network via a stochastic process. In the context of epidemics, there is a related body of research that investigates certain human aspects of decision-making, particularly imitation behavior \citep{fu2011imitation,mbah2012impact}, and empathy \citep{eksin2016disease} in an evolutionary game-theoretic framework. On the other hand, the impacts of human (mis)-perception of probabilities are relatively less explored; an exception is our recent work \cite{hota2018game} which studies game-theoretic choices of curing rates by the nodes in the network.

In this paper, we investigate game-theoretic vaccination decisions against SIS epidemics in general networks under both true as well as human perception of infection probabilities (captured by prospect-theoretic probability weighting functions). Since we consider a cost minimization framework, we refer to players who perceive probabilities as their true values as {\it true expectation minimizers}. We consider the degree-based mean-field (DBMF) approximation \cite{pastor2001epidemic,pastor2015epidemic} to capture the infection probabilities experienced by the nodes. While the DBMF approximation is coarser than the NIMFA, it is more tractable to analyze.

We consider a {\it population game} framework to model strategic decision-making by nodes in large-scale networks; this is motivated by recent works by La on a related class of network security games \cite{la2016interdependent,la2017effects}.\footnote{While La also considers cascades of infection, and some of our structural results such as threshold property and uniqueness of PNE are analogous to those in \cite{la2016interdependent,la2017effects}, our setting with epidemics is quite different from his. In addition, we investigate the impacts of human decision-making while \cite{la2016interdependent,la2017effects} only consider risk neutral decision-makers.} We consider all nodes with a given degree as a single population. The nodes choose whether or not to purchase vaccination at a cost $c > 0$. A vaccinated player is completely immune from the infection. The {\it social state} is defined as the fractions of nodes of each population that vaccinate and remain unprotected. Each node is viewed as an infinitesimal entity, and the change in action of a single node does not change the social state. The social state determines the spreading behavior of the epidemic, which we approximate under the DBMF framework. For a given social state, we identify conditions for the existence and uniqueness of an endemic state of the DBMF approximation where the epidemic persists. The cost of a node who remains unprotected is defined as the perceived infection probability at the endemic state induced by the social state (and therefore, depends on the actions taken by all nodes in the network). 

A pure Nash equilibrium (PNE) of a population game is a social state where no player prefers to unilaterally switch to a different action. We first show that at a PNE social state, all nodes with degrees strictly larger than a threshold vaccinate, and vice versa. In addition, we show that there exists a unique PNE social state. We then prove that when the vaccination is costly, behavioral biases cause fewer nodes to vaccinate, and hence they experience a higher infection probability. In addition, under power-law degree distributions (that are characteristics of real-world networks that arise in different domains \cite{newman2010networks}), we obtain bounds on the equilibrium thresholds for both true and nonlinear perceptions of probabilities, and show that the ratio of these equilibrium thresholds grows unbounded as the vaccination cost increases. We then obtain bounds on the inefficiency of PNE, and numerically illustrate our main theoretical findings.

\section{Nonlinear probability weighting}
\label{section:probweighting}

\begin{figure}[t]
	\begin{center}
	\includegraphics[scale=0.6]{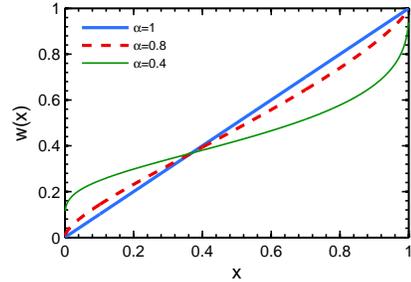}
	\caption{Shape of the probability weighting function \eqref{eq:prelec}; $x$ is the true probability, and $w(x)$ is the perceived probability.}\label{fig:prelecweighting}
	\end{center}
\end{figure}

A substantial body of literature in behavioral economics and psychology has shown that humans perceive probabilities associated with uncertain outcomes in a nonlinear fashion \citep{kahneman1979prospect,dhami2016foundations}. Specifically, humans overweight probabilities that are close to $0$ (referred to as {\it possibility effect}), and underweight probabilities that are close to $1$ (referred to as {\it certainty effect}). In the prospect theory framework \cite{kahneman1979prospect}, Kahneman and Tversky captured the transformation of true probabilities into {\it perceived probabilities} by an inverse S-shaped probability weighting function $w :[0,1] \to [0,1]$ (i.e., a true probability $x$ is perceived as $w(x)$). Several parametric forms of weighting functions have been proposed \cite{tversky1992advances,prelec1998probability}. These weighting functions have the same general shape shown in Figure \ref{fig:prelecweighting}, and satisfy the following properties \cite{hota2016interdependent}. 

\begin{assumption}\label{assumption:weightingfunction}
$w$ satisfies the following properties. 
\begin{enumerate}
\item[1.] $w$ is strictly increasing, with $w(0)=0$ and $w(1)=1$.
\item[2.] \sloppy $w'(x)$ has a unique minimum denoted by $x_{\min,w} := \argmin_{x \in [0,1]} w'(x)$. Furthermore, $w'(x_{\min,w}) < 1$ and $w''(x_{\min,w}) = 0$.
\item[3.] $w(x)$ is strictly concave for $x \in [0,x_{\min,w})$, and is strictly convex for $x \in (x_{\min,w},1]$.
\item[4.] $w'(\epsilon) \to \infty$ as $\epsilon \to 0$, and $w'(1-\epsilon) \to \infty$ as $\epsilon \to 0$.
\end{enumerate}
\end{assumption}
The above assumptions imply that there exists a unique $x_{0,w} \in [0,1]$ such that $w(x) > x$ for $x\in[0,x_{0,w})$, and $w(x) < x$ for $x\in(x_{0,w},1]$.

While our theoretical results hold for any weighting function satisfying Assumption \ref{assumption:weightingfunction}, we consider the parametric form proposed by Prelec \cite{prelec1998probability} in our simulations to give insights on the impacts of smaller and larger deviations from true perception of probability. When the true probability of an outcome is $x$, the Prelec weighting function with parameter $\alpha \in (0,1]$ is given by
\begin{equation}
w(x) = \exp(-(-\ln(x))^\alpha), \text{\qquad} x \in [0,1],
\label{eq:prelec}
\end{equation}
where $\exp(\cdot)$ is the exponential function. For $\alpha=1$, we have $w(x)=x$. It can be verified that for $\alpha \in (0,1)$, the Prelec weighting function \eqref{eq:prelec} satisfies Assumption \ref{assumption:weightingfunction}. For $\alpha < 1$, $w$ has a sharper overweighting of low probabilities and underweighting of high probabilities. Figure \ref{fig:prelecweighting} shows the shape of the Prelec weighting function for different values of $\alpha$. For the Prelec weighting function, $x_{\min,w} = x_{0,w} = \frac{1}{e}$, and $w(\frac{1}{e}) = \frac{1}{e}$ for every $\alpha \in (0,1)$. Furthermore, $w'(\frac{1}{e}) = \alpha$. 
\section{Vaccination Game}
\label{section:vaccination}

We first introduce the population game framework where the nodes decide whether or not to vaccinate themselves. We then discuss how the behavior of the SIS epidemic depends on the decisions taken by the nodes, and define the infection probabilities at the endemic state of the DBMF approximation.

\subsection{Network population game}
\label{subsection:pop_game}

Consider an undirected network $\GG$ where $\DD := \{1,2,\ldots,D\}, D < \infty$, is the set of degrees of the nodes. We treat all nodes with a given degree as a single population, and refer all nodes with degree $d \in \DD$ as population $d$. The fraction of nodes with degree $d$ is denoted by $m_d$. Accordingly, the vector $\{m_d\}_{d\in\DD}$ is the {\it empirical degree distribution} of the network. We denote the average degree by $\langle d\rangle := \sum_{d \in \DD} dm_d$ and define $\langle d^2\rangle := \sum_{d \in \DD} d^2m_d$ with both $\langle d\rangle, \langle d^2\rangle \in (0,\infty)$. We further assume that the network is {\it uncorrelated}, i.e., the probability that an edge originating from a node with degree $k$ is connected to a node with degree $k'$ is independent of $k$. For uncorrelated networks, the probability that a randomly chosen neighbor has degree $d$ is approximately $q_d := \frac{dm_d}{\langle d\rangle}$ \cite{pastor2015epidemic}. The approximation improves as the number of nodes grows to infinity, which is the regime in a population game.

The nodes in the network have two available pure strategies or actions denoted by $\AA = \{\VT,\UT\}$. The state of population $d \in \DD$ is denoted by $\xv_d = (x_{d,\UT},x_{d,\VT})$, where $x_{d,\UT}$ (respectively, $x_{d,\VT}$) represents the subpopulation of nodes that choose to remain unprotected (respectively, vaccinated). Since each node must choose either to remain unprotected or to vaccinate, we have $x_{d,\UT}+x_{d,\VT}=m_d$. Accordingly, the set of feasible population states for population $d$ is $X_d = \{\xv_d \in \Rb_{+}^2 | x_{d,\UT} + x_{d,\VT} = m_d\}$. The {\it social state} is denoted by the vector $\xv = \{\xv_d\}_{d \in \DD} \in X :=\prod_{d \in \DD} X_d$. 

Each node experiences a cost as a function of her degree (i.e., the population she belongs to), her chosen action, and the social state. When the social state is $\xv \in X$, the (expected perceived) cost of a degree $d$ node who chooses an action $a \in \AA$ is denoted by $J_d(a,\xv)$. Following \cite{la2016interdependent}, we define the pure Nash equilibrium (PNE) of the population game as follows. 

\begin{definition}\label{def:pop_game}
A social state $\xv^* \in X$ is a PNE of the population game if for every $a \in \AA, d \in \DD$, we have
\begin{align}\label{eq:def_pne_pop}
x^*_{d,a} > 0 \implies & a \in \argmin_{a' \in \AA} J_{d}(a',\xv^*) \nonumber
\\ \iff & J_{d}(a,\xv^*) \leq J_{d}(a',\xv^*), \forall a' \in \AA.
\end{align}
\end{definition}

According to Definition \ref{def:pop_game}, if a nonzero fraction of nodes of a given population choose a certain action $a \in \AA$ at a PNE, then the action $a$ must achieve the minimum cost for nodes of that population at the PNE.\footnote{In particular, if at a PNE social state $\xv^*$, there is a population $d$ such that both $x^*_{d,\UT} > 0$ and $x^*_{d,\VT} > 0$, then $J_d(\UT,\xv^*) = J_d(\VT,\xv^*)$.} This definition is analogous to that of a PNE in strategic games (with a finite number of players) in the sense that no node prefers to unilaterally deviate to a different action, but the distinction being that such an unilateral deviation does not change the social state in the population game (as opposed to strategic games). Thus, in population games, individual nodes do not have an impact on the population and social states.

The above formulation is motivated by an analogous population game model studied in \cite{la2016interdependent,la2017effects} for a related class of network security games, while in this paper, we focus on the SIS epidemic model. We now describe the dynamics and steady-state of the SIS epidemic for a given social state.

\subsection{SIS epidemic and DBMF approximation}

The spreading process of the SIS epidemic depends on the decisions taken by the players, i.e., the social state of the population game. Specifically, nodes who vaccinate are completely immune from the epidemic, and do not play a role in the spread of the epidemic. On the other hand, each {\it unprotected} node in the network can be in one of two states: i) susceptible, or ii) infected. The {\it vaccinated} nodes are neither susceptible nor infected. Among unprotected nodes, an infected node is cured with a homogeneous {\it curing rate} $\delta > 0$, while a susceptible node becomes infected following a Poisson process with rate $\nu = 1$ (without loss of generality) per infected neighbor. Thus, an unprotected node experiences no risk of infection from its vaccinated neighbors.

Consider a social state $\xv$, which captures the fraction of unprotected and vaccinated nodes in each population.
Under the DBMF approximation \citep{pastor2001epidemic,pastor2015epidemic}, every unprotected node with a given degree $d$ is treated as statistically equivalent.  According to the DBMF approximation, the infection probability of an unprotected degree $d$ node, $\hat{p}_d(\xv,t)$, evolves as
\begin{align}\label{eq:app_dbmf}
\frac{\partial \hat{p}_d(\xv,t)}{\partial t} = -\delta \hat{p}_d(\xv,t) + (1-\hat{p}_d(\xv,t))d \sum_{i \in \mathcal{D}} q_i \frac{x_{i,\UT}}{m_i} \hat{p}_i(\xv,t),
\end{align}
where $q_i = \frac{im_i}{\langle d\rangle}$ is the probability that a randomly chosen neighbor has degree $i$ (as a consequence of the assumption that the network is uncorrelated), and $\frac{x_{i,\UT}}{m_i}$ captures the fraction of degree $i$ nodes that are unprotected in the social state $\xv$. The DBMF approximation is better if the timescale at which nodes interact or mix with each other in the random graph model is faster than the timescale at which the epidemic spreads \cite{pastor2001epidemic,pastor2015epidemic}.

Accordingly, we express the steady-state infection probability of an unprotected degree $d$ node as
\begin{align}
& p_d(\xv) = \frac{d v(\xv)}{\delta + d v(\xv)}, \label{eq:prob_induced}
\end{align}
where $v(\xv)$ denotes the probability that a randomly chosen neighbor is infected in the steady-state of the epidemic induced by the social state $\xv$. In particular, $v(\xv)$ satisfies
\begin{align}
& v(\xv) = \sum_{i \in \DD} q_i \cdot \frac{x_{i,\UT}}{m_i} \cdot \frac{i v(\xv)}{\delta + i v(\xv)}, \nonumber
\\ \implies & v(\xv) \left[ 1- \sum_{i \in \DD} \frac{i \hat{q}_i(\xv)}{\delta + i v(\xv)} \right] = 0, \label{eq:v_induced}
\end{align}
where $\hat{q}_i(\xv) := \frac{ix_{i,\UT}}{\langle d\rangle}$. 

\begin{remark}
The above equation holds under the assumption that the network is uncorrelated. Otherwise, the probability that a randomly chosen neighbor of a degree $d$ node has degree $i$ would depend on both $d$ and $i$, and not just on $i$ as in \eqref{eq:app_dbmf}. Thus, $v(\xv)$ would also be different for different values of $d$. 
\end{remark}

Note that $v(\xv) = 0$ is always a solution of the above equation for every social state, which corresponds to the disease-free state $p_d(\xv)=0, \forall d \in \DD$. Furthermore, depending on the social state $\xv$, there may exist a nonzero $v(\xv) \in (0,1]$ that satisfies \eqref{eq:v_induced}. The set of infection probabilities $p_d(\xv)$ induced by this nonzero $v(\xv)$ according to \eqref{eq:prob_induced} is referred to as the ``endemic" state where the epidemic persists in the population for a long time \cite{van2009virus,pastor2015epidemic}. 

We state the following result on the uniqueness and stability of the endemic state. To the best of our knowledge, the following result is the first time where those properties are shown for the DBMF approximation. While we rely on the results obtained for the NIMFA in \citep{khanafer2016stability,bullo2016lectures}, we exploit certain structural similarities between the two models in the proof. The proof is formally presented in Appendix \ref{appendix_proof}.

\begin{theorem}\label{theorem:app_bullo}
For a given social state $\xv$, define $R(\xv) := \sum_{d\in\mathcal{D}} \frac{d\hat{q}_d(\xv)}{\delta} = \sum_{d\in\mathcal{D}} \frac{d^2 x_{d,\UT}}{\delta \langle d\rangle}$. Then,
\begin{enumerate}
\item $p_d(\xv) = 0, \forall d \in \mathcal{D}$ is the unique steady-state of the dynamics in \eqref{eq:app_dbmf} if and only if $R(\xv) \leq 1$. This disease free steady-state is globally asymptotically stable.
\item If $R(\xv) > 1$, $p_d(\xv) = 0, \forall d \in \mathcal{D}$ is an unstable equilibrium of the dynamics \eqref{eq:app_dbmf}. There exists a nonzero positive steady-state of \eqref{eq:app_dbmf} (referred to as the endemic state) with $p_d(\xv) > 0, \forall d \in \mathcal{D}$ if and only if $R(\xv) > 1$. The endemic state is unique, locally exponentially stable, and the dynamics \eqref{eq:app_dbmf} converge to this endemic state from any initial condition except the disease free state. 
\end{enumerate}
\end{theorem}

It follows from the above theorem that if $R(\xv) > 1$ for a social state $\xv$, then there exists a unique nonzero $v(\xv)$ satisfying \eqref{eq:v_induced}. In that case, we interpret $v(\xv)$ to be this nonzero solution. The infection probabilities of the unprotected nodes in the endemic state are then nonzero and given by \eqref{eq:prob_induced}. 

\section{Properties of Pure Nash Equilibrium}\label{section:nash}

With the above framework in place, we now define the cost functions of the players and establish the existence of a unique threshold equilibrium for both true and perceived expectation minimizers in general networks. 

Each node independently decides whether or not to vaccinate. For a node who remains unprotected, we define her cost as her perceived infection probability at the endemic state under the DBMF approximation of the epidemic. Specifically, the cost of an unprotected degree $d$ node in social state $\xv$ is 
\begin{equation}\label{eq:cost}
J_d(\UT,\xv) = w(p_d(\xv)) = w\left( \frac{d v(\xv)}{\delta + d v(\xv)} \right),
\end{equation}
where $w$ is the probability weighting function of the nodes, and $v(\xv)$ satisfies \eqref{eq:v_induced}. Specifically, if $R(\xv) \leq 1$, $v(\xv) = 0$. Otherwise, $v(\xv)$ is the nonzero solution of \eqref{eq:v_induced}. The above cost function can be interpreted as the expected cost under cumulative prospect theory \cite{tversky1992advances}, where the perceived loss of being infected is normalized to $1$.

On the other hand, we assume that vaccination is available at a cost $c \in (0,1)$ irrespective of the degrees of the nodes. Therefore, $J_d(\VT,\xv) = c$ for every $d \in \DD, \xv \in X$. Here $c$ can be interpreted as the relative cost of vaccination compared to the cost of being infected. Note that if $c \geq 1$, we will always have $J_d(\UT,\xv) < c$, i.e., all nodes will prefer to remain unprotected. 

We consider a complete information game denoted by $\Gamma(\GG,\{m_d\}_{d \in \DD},w,c,\delta)$. Since $\Gamma(\GG,\{m_d\}_{d \in \DD},w,c,\delta)$ has a finite number of {\it populations} and {\it actions}, there always exists a pure Nash equilibrium (PNE) \cite{sandholm2010population}.

\begin{remark}
As in any complete information game, the nodes are assumed to be aware of the social state and the degree distribution, from which they can compute the quantity $v(x)$. We envision two possible ways in which nodes can estimate $v(x)$ in practice. First, a central authority broadcasts this information among the population; for instance, via news and social media. Second, nodes can learn or estimate $v(x)$ by repeatedly interacting with their neighbors, and observing the frequency with which they get infected. It would be an interesting direction for future research to explicitly model the dynamics of the epidemic game where nodes estimate $v(x)$ and update their vaccination decisions, and explore whether the dynamics converge to the equilibria studied in this work.
\end{remark}

We now discuss the characteristics of vaccination strategies that arise at a Nash equilibrium. We start with the following result which shows that if the curing rate is not sufficiently high, then at every PNE social state, there exists an endemic state such that the epidemic persists. 

\begin{proposition}
Let $\xv^*$ be the social state at a PNE. Then, $R(\xv^*) > 1$ if and only if $\delta < \frac{\langle d^2\rangle}{\langle d\rangle} = \frac{\sum_{d \in \DD} d^2m_d}{\sum_{d \in \DD} dm_d}$.
\end{proposition}
\begin{IEEEproof}
Let $\delta < \frac{\langle d^2\rangle}{\langle d\rangle}$. Assume on the contrary that $R(\xv^*) \leq 1$ for a PNE social state $\xv^*$. Then, from Theorem \ref{theorem:app_bullo}, we have $v(\xv^*) = 0$. Now, let $d \in \DD$ be a population of nodes such that a non-negligible fraction of degree $d$ nodes vaccinate, i.e., $x^*_{d,\VT} > 0$. However, following Definition \ref{def:pop_game}, this implies $c \leq w(p_d(\xv^*)) = 0$. Since $c > 0$, we must necessarily have $x^*_{d,\VT} = 0$, or equivalently, $x^*_{d,\UT} = m_d$ for every $d \in \DD$. As a consequence, we have
$$ R(\xv^*) = \sum_{d \in \DD} \frac{d^2 m_d}{\delta \langle d\rangle} = \frac{\langle d^2\rangle}{\delta \langle d\rangle} > 1,$$
which is the desired contradiction. 

For the converse, we need to show that $R(\xv^*) > 1 \implies \delta < \frac{\langle d^2\rangle}{\langle d\rangle}$. We prove the statement by contraposition. Suppose $\delta \geq \frac{\langle d^2\rangle}{\langle d\rangle}$. Now, consider a PNE social state $\xv^*$. Then, 
$$ R(\xv^*) = \sum_{d \in \DD} \frac{d^2 x^*_{d,\UT}}{\delta \langle d\rangle} \leq \frac{\langle d^2\rangle}{\delta \langle d\rangle} \leq 1.$$
This completes the proof.
\end{IEEEproof}

The above result holds for both true as well as perceived expected cost minimizers. The proof shows that if $\delta \geq \frac{\langle d^2\rangle}{\langle d\rangle}$, then even for a social state where no node purchases vaccination, the disease free state is the only solution of the DBMF approximation. In other words, the epidemic disappears solely because of high curing rate, and independently of the vaccination decisions by the players. Therefore, we focus on the more interesting case where the spread of the disease depends on the vaccination decisions of the nodes. 

Note that a similar result was obtained in \cite{omic2009protecting} where nodes chose their curing rates strategically. At the resulting equilibrium, the quantity analogous to $R(\xv^*)$ was greater than or equal to $1$, but never strictly less than $1$. We now impose the following assumption in the rest of the paper.

\begin{assumption}\label{assumption:curing}
The curing rate $\delta \in \left(0,\frac{\langle d^2\rangle}{\langle d\rangle}\right)$. Furthermore, either $w(x)=x$, or $w$ satisfies Assumption \ref{assumption:weightingfunction}.
\end{assumption} 

We now derive certain properties of the social states at the PNE. Let $\xv^*$ be the social state at a PNE. Consider a degree $d$ such that $x^*_{d,\UT} > 0$, i.e., a nonzero fraction of degree $d$ nodes have not purchased vaccination. Then, we must have
\begin{align}\label{eq:PNE_leq}
w(p_d(\xv^*)) & = w\left(\frac{dv(\xv^*)}{\delta+dv(\xv^*)}\right) \leq c.
\end{align}
Similarly, if for the population of degree $d$ nodes $x^*_{d,\VT} = m_d - x^*_{d,\UT} > 0$, we must have
\begin{align}\label{eq:PNE_geq}
w(p_d(\xv^*)) & = w\left(\frac{dv(\xv^*)}{\delta+dv(\xv^*)}\right) \geq c.
\end{align}

The following result shows that any PNE of the vaccination game exhibits a threshold property. 

\begin{proposition}\label{proposition:NE_threshold}
Let $\xv^*$ be the social state at a PNE. Then, there exists a degree $\underline{d}$ such that $x^*_{\underline{d},\UT} \in (0,m_{\underline{d}}]$, and 
\begin{equation}\label{eq:thresholdNE_def}
x^*_{d,\UT} = 
\begin{cases}
m_d, & \quad d < \underline{d}
\\ 0, & \quad d > \underline{d}.
\end{cases}
\end{equation}
\end{proposition}
\begin{IEEEproof}
Let $\bar{d} \in \DD$ be the smallest degree such that a nonzero fraction of nodes with degree $\bar{d}$ vaccinate, i.e., $m_{\bar{d}} - x^*_{\bar{d},\UT} > 0$. Since $w$ is an increasing function, we have
$$ c \leq w\left(\frac{\bar{d}v(\xv^*)}{\delta+\bar{d}v(\xv^*)}\right) < w\left(\frac{dv(\xv^*)}{\delta+dv(\xv^*)}\right), $$
for every $d > \bar{d}$. Note that $v(\xv^*) > 0$ following Assumption \ref{assumption:curing}. Accordingly, $x^*_{d,\UT} = 0$ for every $d > \bar{d}$. In other words, all nodes with degree strictly larger than $\bar{d}$ vaccinate. We now have the following two possible cases. 

\noindent {\bf Case 1: $x^*_{\bar{d},\UT} = 0$}. In this case, all nodes with degree $\bar{d}$ vaccinate. Since $\bar{d}$ is the smallest degree with a nonzero fraction of vaccinated nodes, all nodes with degree $d < \bar{d}$ remain unprotected. Therefore, $\underline{d} := \bar{d}-1$ has the threshold property stated in \eqref{eq:thresholdNE_def} (with $x^*_{\underline{d},\UT} = m_{\underline{d}}$).

\noindent {\bf Case 2: $x^*_{\bar{d},\UT} > 0$}. In this case, we have
$$ c = w\left(\frac{\bar{d}v(\xv^*)}{\delta+\bar{d}v(\xv^*)}\right) > w\left(\frac{dv(\xv^*)}{\delta+dv(\xv^*)}\right), $$
for all $d < \bar{d}$. Therefore, $x^*_{d,\UT} = m_d$ for every $d < \underline{d}$. Thus, $\underline{d} := \bar{d}$ has the threshold property stated in \eqref{eq:thresholdNE_def}.
\end{IEEEproof}

Note that high degree nodes are more likely to encounter infected nodes among their neighbors than nodes with a smaller degree. This is because the network is assumed to be uncorrelated. The above result then confirms our intuition that high degree nodes are more likely to vaccinate at the equilibrium as they experience a higher infection risk. 

A similar threshold behavior was also obtained in \cite{trajanovski2015decentralized} under the NIMFA for specific classes of networks (specifically, complete graphs, complete bipartite graphs and multi-community networks). Our result shows that this property holds under the DBMF approximation in general networks. Furthermore, in \cite{pastor2002immunization}, it was shown that in networks with power-law degree distributions, vaccinating high degree nodes is highly effective in reducing the spread of the epidemic. In this paper, we show that such strategies arise naturally under decentralized decision-making, and in networks that are uncorrelated (as opposed to only under power-law degree distributions).

Following Proposition \ref{proposition:NE_threshold}, we observe that the population states of unprotected nodes at any PNE are of the form
\begin{equation}\label{eq:candidate_popstate}
\xv^*_{\UT} =\{m_1,m_2,\ldots,m_{\underline{d}-1},x^*_{\underline{d},\UT},0,\ldots,0\},
\end{equation}
with $x^*_{\underline{d},\UT} \in (0,m_{\underline{d}}]$ for some $\underline{d}$. Therefore, we refer to social states where the population states of unprotected nodes have the above form as {\it candidate social states}. We define the following preference relation between candidate social states. 

\begin{definition}\label{def:pref_socstate}
For two candidate social states $\xv_1$ and $\xv_2$ with populations of unprotected nodes given by
\begin{align}
\xv_{1,\UT} & = \{m_1,m_2,\ldots,m_{\underline{d}_1-1}, x_{1,\underline{d}_1,\UT}, 0, \ldots, 0\}, \quad \text{and} \label{eq:cand_pop_1}
\\ \xv_{2,\UT} & = \{m_1,m_2,\ldots,m_{\underline{d}_2-1}, x_{2,\underline{d}_2,\UT}, 0, \ldots, 0 \}, \label{eq:cand_pop_2}
\end{align}
we say $\xv_1 \prec \xv_2$ if (i) either $\underline{d}_1 < \underline{d}_2$ or (ii) $\underline{d}_1 = \underline{d}_2$ and $x_{1,\underline{d}_1,\UT} < x_{2,\underline{d}_2,\UT}$. Similarly, we say $\xv_1 \preceq \xv_2$ if either $\xv_1 \prec \xv_2$ or $\underline{d}_1 = \underline{d}_2$ and $x_{1,\underline{d}_1,\UT} = x_{2,\underline{d}_2,\UT}$. 
\end{definition}

In other words, $\xv_1 \preceq \xv_2$ if the fraction of unprotected nodes in $\xv_1$ is at most that of $\xv_2$. We now state the following lemma which will be useful in proving several of our results. 

\begin{lemma}\label{lemma:v_monotone}
Consider two candidate social states $\xv_1$ and $\xv_2$. If $\xv_1 \preceq \xv_2$, then $v(\xv_1) \leq v(\xv_2)$ with the inequality being strict if $\xv_1 \prec \xv_2$.
\end{lemma}
\begin{IEEEproof}
Suppose $v(\xv_1) > 0$; otherwise the result trivially holds. If $\xv_1 \preceq \xv_2$, from \eqref{eq:v_induced} we have
$$\sum_{d \in \DD} \frac{d^2 x_{1,d,\UT}}{\delta + d v(\xv_1)} = \langle d\rangle = \sum_{d \in \DD} \frac{d^2 x_{2,d,\UT}}{\delta + d v(\xv_2)} \geq \sum_{d \in \DD} \frac{d^2 x_{1,d,\UT}}{\delta + d v(\xv_2)}. $$
The above equation implies $v(\xv_1) \leq v(\xv_2)$. Furthermore, the inequality in the above equation is strict if $\xv_1 \prec \xv_2$.
\end{IEEEproof}

The above lemma shows that among two candidate social states, if one of the social states has a larger fraction of unprotected nodes, then for that social state, the probability of a randomly chosen neighbor being infected is higher. The above lemma and the following uniqueness result hold for both true and nonlinear perception of probabilities.

\begin{proposition}\label{proposition:unique_v}
Let $\xv_1$ and $\xv_2$ be two PNE social states. Then, $v(\xv_1) = v(\xv_2)$. Furthermore, $\xv_1 = \xv_2$.
\end{proposition}
\begin{IEEEproof}
Following Proposition \ref{proposition:NE_threshold}, let $\underline{d}_1$ and $\underline{d}_2$ denote the thresholds for social states $\xv_1$ and $\xv_2$, respectively. In other words, $\underline{d}_1$ and $\underline{d}_2$ are the largest degrees such that $x_{1,\underline{d}_1,\UT} > 0$ and $x_{2,\underline{d}_2,\UT} > 0$, respectively. Furthermore, the population of unprotected nodes in $\xv_1$ and $\xv_2$ can be represented as \eqref{eq:cand_pop_1} and \eqref{eq:cand_pop_2}, respectively. Note that $x_{2,\underline{d}_2,\UT} \leq m_{\underline{d}_2}$. Accordingly, from \eqref{eq:v_induced}, $v(\xv_2)$ satisfies
\begin{align}\label{eq:v_unique_internal2}
1 & = \sum_{d \in \DD} \frac{d \hat{q}_d(\xv_2)}{\delta + d v(\xv_2)} \leq \frac{1}{\langle d\rangle} \sum^{\underline{d}_2}_{d =1} \frac{d^2m_d}{\delta + d v(\xv_2)}.
\end{align}

Now assume on the contrary that $v(\xv_1) < v(\xv_2)$ without loss of generality. From the definition of PNE, we have
\begin{align}
& w\left(\frac{\underline{d}_2 v(\xv_1)}{\delta+ \underline{d}_2 v(\xv_1)}\right) < w\left(\frac{\underline{d}_2 v(\xv_2)}{\delta+ \underline{d}_2 v(\xv_2)}\right) \leq c \label{eq:v_unique_internal}
\\ \implies & x_{1,\underline{d}_2,\UT} = m_{\underline{d}_2}. \nonumber
\end{align}
In other words, all $\underline{d}_2$ nodes remain unprotected at $\xv_1$. Accordingly, we have $\underline{d}_2 \leq \underline{d}_1$. Thus, $v(\xv_1)$ satisfies
\begin{align}\label{eq:v_unique_internal3}
1 & = \sum_{d \in \DD} \frac{d \hat{q}_d(\xv_1)}{\delta + d v(\xv_1)} \geq \frac{1}{\langle d\rangle} \sum^{\underline{d}_2}_{d =1} \frac{d^2m_d}{\delta + d v(\xv_1)};
\end{align}
the above inequality is strict if $\underline{d}_2 < \underline{d}_1$, and also holds when $\underline{d}_2 = \underline{d}_1$ since $x_{1,\underline{d}_2,\UT} = m_{\underline{d}_2}$. 

Therefore, from equations \eqref{eq:v_unique_internal2} and \eqref{eq:v_unique_internal3}, we have
$$ \sum^{\underline{d}_2}_{d =1} \frac{d^2m_d}{\delta + d v(\xv_1)} \leq \sum^{\underline{d}_2}_{d =1} \frac{d^2m_d}{\delta + d v(\xv_2)} \implies v(\xv_1) \geq v(\xv_2),$$
which gives the desired contradiction.

Consequently, we have $v(\xv_1) = v(\xv_2)$. Now, suppose $\xv_1 \prec \xv_2$. Then, Lemma \ref{lemma:v_monotone} implies $v(\xv_1) < v(\xv_2)$, which is a contradiction. Thus, we must have $\xv_1 = \xv_2$. 
\end{IEEEproof} 

\section{Comparison of Equilibria under True and Perceived Expectation Minimizers}
\label{section:comparison}

Having established the existence and uniqueness of PNE in the vaccination game under both true and perceived expectation minimizers, we now compare their characteristics. 

Let $\xv^w$ be the PNE social state, and $\xv^w_{\UT}$ be the population of unprotected nodes at the PNE of a game where the probability weighting function of the nodes is $w$. Following Proposition \ref{proposition:NE_threshold}, $\xv^w_{\UT} = \{m_1,m_2,\ldots,x^w_{d_w},0,\ldots,0\}$, where $x^w_{d_w} \in (0,m_{d_w}]$. Let $v(\xv^w)$ be the probability that a randomly chosen neighbor is infected at the PNE. From the definition of PNE, we have
\begin{equation}\label{eq:pne_weighting}
w\left(\frac{d_w v(\xv^w)}{\delta+d_w v(\xv^w)} \right) \leq c \leq w\left(\frac{(d_w+1) v(\xv^w)}{\delta+(d_w+1) v(\xv^w)} \right),
\end{equation}
with at least one of the inequalities being strict. Since $w$ is strictly increasing, from the first inequality above, we have
\begin{align*}
& \frac{d_w v(\xv^w)}{\delta+d_w v(\xv^w)} \leq w^{-1}(c) 
\\ \implies & 1 - w^{-1}(c) \leq 1 - \frac{d_w v(\xv^w)}{\delta+d_w v(\xv^w)} = \frac{\delta}{\delta+d_w v(\xv^w)}
\\ \implies & d_w v(\xv^w) \leq \frac{\delta w^{-1}(c)}{1-w^{-1}(c)}.
\end{align*}
Following identical steps for the second inequality, we conclude that \eqref{eq:pne_weighting} is equivalent to
\begin{equation}\label{eq:pne_weighting_eq}
d_w v(\xv^w) \leq \frac{\delta w^{-1}(c)}{1-w^{-1}(c)} \leq (d_w+1) v(\xv^w).
\end{equation}

We now state the following main result.
\begin{proposition}\label{proposition:threshold_comparison}
Let $\xv^t$ be the PNE social state under true expectation minimizers. If the  vaccination cost $c$ satisfies $c \geq w(c)$, then we have $\xv^t \preceq \xv^w$, and vice versa.
\end{proposition}
\begin{IEEEproof}
Let $\xv^t_{\UT} = \{m_1,m_2,\ldots,x^t_{d_t},0,\ldots,0\}$. Suppose $c \geq w(c)$, and assume on the contrary that $\xv^w \prec \xv^t$. Following Lemma \ref{lemma:v_monotone}, we have $v(\xv^t) > v(\xv^w)$. Therefore, we have
$$ d_tv(\xv^w) < d_tv(\xv^t) \leq \frac{\delta c}{1-c} \leq \frac{\delta  w^{-1}(c)}{1- w^{-1}(c)}, $$
where the second inequality follows from \eqref{eq:pne_weighting_eq}, and the last inequality follows since $c \geq w(c)$, or equivalently, $w^{-1}(c) \geq c$. Consequently, $d_tv(\xv^w) < \frac{\delta  w^{-1}(c)}{1- w^{-1}(c)}$ implies
\begin{align*}
&  d_tv(\xv^w)(1- w^{-1}(c)) < \delta  w^{-1}(c)
\\ \implies & w\left( \frac{d_tv(\xv^w)}{\delta+d_tv(\xv^w)} \right) < c \implies x^w_{d_t} = m_{d_t},
\end{align*}
i.e., all nodes of degree $d_t$ are unprotected at the PNE under probability weighting. Therefore, we have $\xv^t \preceq \xv^w$, which is the desired contradiction. The case where $c \leq w(c)$ follows from identical arguments, and in that case, we have $\xv^w \preceq \xv^t$. We omit the details in the interest of space.
\end{IEEEproof}

The key intuition behind the above result is that the vaccination cost is very close to the perceived infection probability at the PNE threshold, which is a consequence of \eqref{eq:pne_weighting}. Consequently, at a high vaccination cost, the perceived infection probability at the equilibrium threshold is also high. For a given perceived probability, the true probability is higher under nonlinear probability weighting compared to true expectation minimizers, due to underweighting of high probabilities. Thus, for large vaccination costs, behavioral biases cause fewer nodes to vaccinate compared to the case with true expectation minimizers. Specifically, the equilibrium threshold is higher under nonlinear probability weighting. The converse is true when the vaccination cost is sufficiently small where the decision-makers overweight low infection probabilities. Consequently, for low vaccination costs, a higher population of nodes vaccinate at the PNE under probability weighting.  

We emphasize that the above result holds irrespective of the degree distribution of the network, and only depends on the vaccination cost $c$ and the probability weighting function $w$. The generality of this result has significant policy implications: it is beneficial to reduce vaccination costs below the level where $w(c)=c$ such that behavioral biases, specifically overweighting of low probabilities, lead to a higher rate of vaccination. In contrast, if behavioral biases are not taken into account and vaccines are relatively costly, vaccination rates among the human population could be far lower than what is anticipated for risk neutral decision makers. The following subsection quantifies this effect in a class of networks with a power-law degree distribution.


\subsection{Networks with a power-law degree distribution}

We now consider a class of networks where the set of degrees is $\DD =\{d_0, d_1, \ldots, D\}$ with $d_0 \geq 1, D < \infty$. We assume that the fraction of nodes with degree $d \in \DD$ obeys the power-law with exponent $\beta \in [2,3]$, i.e., $m_d = \kappa d^{-\beta}$, where $\kappa = (\sum_{d \in \DD} d^{-\beta})^{-1}$ is the normalization constant. Our primary motivation for considering this model is that networks with a power-law degree distribution with exponent $3$ (and more generally in the interval $[2,3]$) arise in a wide range of settings, and in particular, under the the Barab{\'a}si-Albert (BA) preferential attachment model \cite{newman2010networks}.

Before comparing the equilibrium thresholds, we first obtain bounds on the steady-state infection probability for threshold-based social states. Let $\xv_t$ be the social state and $\xv_{t,\UT}$ be the population of unprotected nodes where all nodes with degree at most $t$ are unprotected, and all nodes with degree strictly larger than $t$ are vaccinated, i.e., 
\begin{equation}\label{eq:th_popstate}
\xv_{t,\UT} := \{m_{d_0},m_{d_1},\ldots,m_{t},0,\ldots,0\}.
\end{equation}
The probability that a randomly chosen neighbor is infected at the endemic state under this social state is denoted by $v_{\beta}(\xv_t)$ for a power-law distribution with exponent $\beta$; $v_{\beta}(\xv_t)$ satisfies
\begin{equation}\label{eq:th_v}
\!1 \!= \!\sum^t_{i = d_0} \!\!\frac{i^2m_i}{\langle d\rangle(\delta+iv_{\beta}(\xv_t))} \!= \!\frac{\kappa}{\langle d\rangle} \sum^t_{i = d_0} \!\!\frac{1}{i^{\beta-2}(\delta+iv_{\beta}(\xv_t))}.
\end{equation}

We first obtain the following bounds.

\begin{lemma}\label{lemma:th_bound}
Consider a network with a power-law degree distribution with exponent $\beta \in [2,3]$. Let the threshold $t$ be large enough such that $v_{\beta}(\xv_t) > 0$. Then,
\begin{equation}\label{eq:th_lower}
\frac{t-d_0B_1}{t-d_0} \leq \frac{tv_{\beta}(\xv_t)}{\delta+tv_{\beta}(\xv_t)},
\end{equation}
where $B_1 = \exp(\frac{\delta \langle d\rangle}{\kappa})$. If, in addition, $\beta = 3$ and $d_0 > 1$, then
\begin{equation}\label{eq:th_upper}
\frac{tv_3(\xv_t)}{\delta+tv_3(\xv_t)} \leq \frac{t-(d_0-1)B_1}{t-d_0+1}.
\end{equation}
\end{lemma}

The result is obtained by bounding the summation in \eqref{eq:th_v} with appropriate integrals. The proof is presented in Appendix \ref{appendix:lemma}. We now apply the above bounds to show how equilibrium thresholds behave as functions of the vaccination cost $c$. 

\begin{proposition}\label{proposition:powerlaw_rev}
Consider a vaccination game on a network with a power-law degree distribution with exponent $\beta \in [2,3]$. Let $d_w(c)$ denote the equilibrium threshold when the vaccination cost is $c$, and the weighting function is $w$. Then, 
$$d_w(c) \leq \min\left(D,1 + d_0 + \frac{d_0(B_1-1)}{1-w^{-1}(c)}\right).$$
\end{proposition}
\begin{IEEEproof}
Let $\xv^w(c)$ denote the PNE social state. Suppose $d_w(c) \geq d_0+1$; otherwise the result trivially holds. Recall from \eqref{eq:th_popstate} that a social state where all nodes up to degree $d_w(c)-1$ are unprotected is denoted by $\xv_{d_w(c)-1}$. Following Lemma \ref{lemma:v_monotone}, we have $v(\xv^w(c)) > v(\xv_{d_w(c)-1})$. From the definition of PNE and \eqref{eq:th_lower} in Lemma \ref{lemma:th_bound}, we obtain
\begin{align*}
& c \geq w\left(\frac{(d_w(c)-1)v(\xv_{d_w(c)-1})}{\delta+(d_w(c)-1)v(\xv_{d_w(c)-1})}\right) 
\\ \implies & w^{-1}(c) \geq \frac{d_w(c)-1-d_0B_1}{d_w(c)-1-d_0}
\\ \implies & 1 - w^{-1}(c) \leq 1 - \frac{d_w(c)-1-d_0B_1}{d_w(c)-1-d_0} = \frac{d_0(B_1-1)}{d_w(c)-1-d_0}
\\ \implies & d_w(c)-1-d_0 \leq \frac{d_0(B_1-1)}{1-w^{-1}(c)}.
\end{align*}
The result holds as the maximum degree is $D$.
\end{IEEEproof}

Note that the upper bound obtained above does not depend on the power-law exponent $\beta$ as long as $\beta \in [2,3]$. Furthermore, the bound on the equilibrium vaccination threshold increases as the vaccination cost $c$ increases to $1$, and is inversely proportional to $1-w^{-1}(c)$. In particular, as $c \to 1$, underweighting of high probabilities implies that $w^{-1}(c) > c$. Accordingly, $1-c > 1-w^{-1}(c)$, and the bound on the threshold is higher under probability weighting, which is consistent with our observation from Proposition \ref{proposition:threshold_comparison}.

In the special case where the exponent $\beta=3$, the following result obtains tight bounds on the ratio of equilibrium thresholds under nonlinear and true perceptions of probabilities.

\begin{proposition}\label{proposition:powerlaw3}
Consider a network with a power-law degree distribution with exponent $\beta=3$ and $d_0 > 1$. For vaccination cost $c$, denote equilibrium thresholds under true and perceived expectation minimizers to be $d_t(c)$ and $d_w(c)$. Then, 
$$\frac{d_w(c)}{d_t(c)} = \Theta\left(\frac{1-c}{1-w^{-1}(c)}\right).$$
\end{proposition}

\begin{IEEEproof}
First, we consider the PNE under true expectation minimizers. Following Proposition \ref{proposition:powerlaw_rev}, we have
$$d_t(c)-1-d_0 \leq \frac{d_0(B_1-1)}{1-c}.$$
We now obtain a lower bound on $d_t(c)$ following the second part of Lemma \ref{lemma:th_bound}. Let $\xv^t(c)$ be the social state at the PNE with true expectation minimizers. Since all nodes with degree at most $d_t(c)$ are unprotected at the social state $\xv_{d_t(c)}$, we have $\xv^t(c) \preceq \xv_{d_t(c)}$, and following Lemma \ref{lemma:v_monotone}, $v(\xv^t(c)) \leq v(\xv_{d_t(c)})$. From the definition of the PNE and together with \eqref{eq:th_upper} in Lemma \ref{lemma:th_bound}, we obtain
\begin{align*}
& c \leq \frac{d_t(c) v(\xv_{d_t(c)})}{\delta+ d_t(c) v(\xv_{d_t(c)})} \leq \frac{d_t(c)-(d_0-1)B_1}{d_t(c)-d_0+1}
\\ \implies & 1-c \geq 1 - \frac{d_t(c)-(d_0-1)B_1}{d_t(c)-d_0+1} = \frac{(d_0-1)(B_1-1)}{d_t(c)+1-d_0}
\\ \implies & d_t(c)+1-d_0 \geq \frac{(d_0-1)(B_1-1)}{1-c}.
\end{align*}
From the above computations, we have $d_t(c) = \Theta((1-c)^{-1})$. Following the definition of PNE under nonlinear probability weighting \eqref{eq:pne_weighting}, and repeating the above arguments, we have $d_w(c) = \Theta((1-w^{-1}(c))^{-1})$. This concludes the proof.
\end{IEEEproof}

Thus, the upper bound in \eqref{eq:th_upper} in Lemma \ref{lemma:th_bound} enables us to derive a more refined result for the special case when the exponent $\beta=3$. Recall from Assumption \ref{assumption:weightingfunction} that $w'(1-\epsilon) \to \infty$ as $\epsilon \to 0$. Therefore, as $c \to 1$, $\left(\frac{1-c}{1-w^{-1}(c)}\right) \to \infty$. Thus, in networks with a power-law degree distribution with exponent $3$, the threshold under probability weighting is significantly higher compared to the threshold under true expectation minimizers when $c \to 1$. In other words, when the vaccination cost approaches the cost of being infected, underweighting of probabilities causes even nodes with very high degrees to not vaccinate against the epidemic. As a result, the network could be highly susceptible to the spread of the epidemic. Our results highlight the importance of incorporating the behavioral biases of the decision-makers; otherwise we could significantly underestimate the epidemic risks in human populations.
\section{Social Optimum and Inefficiency of PNE}
\label{section:nash}

In order to characterize the inefficiency of the PNE, we consider the case where a central authority decides which nodes to vaccinate (i.e., the social state) in order to minimize a social cost. We start by showing that the socially optimal vaccination policy also has a threshold property similar to that at a PNE. We define the social cost at a social state $\xv$ as
\begin{align}
\Psi(\xv) & := \sum_{d \in \DD} \sum_{a \in \AA} x_{d,a} J_d(a,\xv) \nonumber
\\ & = \sum_{d \in \DD} \left[x_{d,\UT} \frac{dv(\xv)}{\delta+dv(\xv)} + (m_d - x_{d,\UT}) c\right] \nonumber
\\ & = c + \sum_{d \in \DD} x_{d,\UT} \left[\frac{dv(\xv)}{\delta+dv(\xv)} -c\right], \label{eq:soc_cost}
\end{align}
since $\sum_{d \in \DD} m_d = 1$. Note from the second equality above that the social cost is composed of two terms; the first is the expected fraction of nodes that are infected in the endemic state and the second is normalized cost of vaccination. Furthermore, the social cost is the total {\it true} expected cost of all populations, i.e., the central authority does not have behavioral biases. The following result shows that the socially optimal vaccination policy follows a threshold behavior.

\begin{proposition}\label{proposition:th_opt}
Let $\yv^* \in \argmin_{\xv \in X} \Psi(\xv)$. Suppose there exists a degree $d'$ such that $y^*_{d',\UT} < m_d$. Then, $y^*_{d,\UT} = 0$ for every $d > d'$.
\end{proposition}
\begin{IEEEproof}
Note that $\Psi(\xv)$ is continuous in $\xv$; when $v(\xv)$ is nonzero, it is continuous in $\xv$ following \eqref{eq:v_induced}. Thus, $\yv^*$ exists. Assume on the contrary that there exist degrees $d_1$ and $d_2$, with $d_1 < d_2$ such that $y^*_{d_1,\UT} < m_{d_1}$ and $y^*_{d_2,\UT} > 0$. We now construct a social state $\bar{\yv}$ with a smaller social cost. Let $0 < \epsilon < \min(y^*_{d_2,\UT},m_{d_1} - y^*_{d_1,\UT})$. We define
\begin{equation}\label{eq:bary}
\bar{y}_{d,\UT} =   
\begin{cases} 
      \hfill y^*_{d_1,\UT} + \epsilon \hfill & \text{if $d=d_1$} \\
      \hfill y^*_{d_2,\UT} - \epsilon \hfill & \text{if $d=d_2$}, \\
      \hfill y^*_{d,\UT} \hfill & \text{otherwise.}
\end{cases}
\end{equation}
In other words, in $\bar{\yv}$, a smaller fraction of nodes with degree $d_1$ are vaccinated, and a larger fraction of nodes with degree $d_2$ are vaccinated compared to $\yv^*$. 

Before comparing the social costs at $\yv^*$ and $\bar{\yv}$, we first claim that $v(\bar{\yv}) < v(\yv^*)$. Suppose not, and we instead have $v(\bar{\yv}) \geq v(\yv^*)$. From \eqref{eq:v_induced}, we have
\begin{align*}
\langle d\rangle = & \sum_{d \in \DD} \frac{d^2 y^*_{d,\UT}}{\delta + d v(\yv^*)} = \sum_{d \in \DD} \frac{d^2 \bar{y}_{d,\UT}}{\delta + d v(\bar{\yv})} \leq \sum_{d \in \DD} \frac{d^2 \bar{y}_{d,\UT}}{\delta + d v(\yv^*)}
\\ \implies & \sum_{d \in \DD} \frac{d^2 (y^*_{d,\UT} - \bar{y}_{d,\UT})}{\delta + d v(\yv^*)} \leq 0
\\ \implies &  \frac{-\epsilon d_1^2}{\delta + d_1 v(\yv^*)} + \frac{\epsilon d_2^2}{\delta + d_2 v(\yv^*)} \leq 0, \qquad \qquad \text{from \eqref{eq:bary}}
\\ \implies & d^2_2 (\delta+d_1v(\yv^*)) \leq d^2_1 (\delta+d_2v(\yv^*))
\\ \implies & \delta(d^2_2-d^2_1) + d_1d_2v(\yv^*)(d_2-d_1) \leq 0,
\end{align*}
which is a contradiction, since $d_2 > d_1$ and the left hand side is strictly positive. Thus, we have $v(\bar{\yv}) < v(\yv^*)$. 

We now compute the difference $\Psi(\bar{\yv}) - \Psi(\yv^*)$ as
\begin{align*}
& \sum_{d \in \DD} \bar{y}_{d,\UT} \left[\frac{dv(\bar{\yv})}{\delta+dv(\bar{\yv})} -c\right] - \sum_{d \in \DD} y^*_{d,\UT} \left[\frac{dv(\yv^*)}{\delta+dv(\yv^*)} -c\right] 
\\ & < \sum_{d \in \DD} (\bar{y}_{d,\UT} - y^*_{d,\UT})\frac{dv(\yv^*)}{\delta+dv(\yv^*)}
\\ & = \epsilon \left[\frac{d_1v(\yv^*)}{\delta+d_1v(\yv^*)} - \frac{d_2v(\yv^*)}{\delta+d_2v(\yv^*)}\right] < 0,
\end{align*}
where the first inequality is a consequence of $v(\bar{\yv}) < v(\yv^*)$ and $\sum_{d \in \DD} \bar{y}_{d,\UT} = \sum_{d \in \DD} y^*_{d,\UT}$ (following \eqref{eq:bary}). The second inequality holds since $d_2 > d_1$. This contradicts the claim that $\yv^*$ is the socially optimal social state.
\end{IEEEproof}

The above result shows that a central authority would also choose to vaccinate high degree nodes, and keep nodes with a smaller degree unprotected. Furthermore, the population states of unprotected nodes at a social optimum have a similar structure as \eqref{eq:candidate_popstate}. Despite this structural similarity with the PNE vis-a-vis Proposition \ref{proposition:NE_threshold}, we show that a larger fraction of nodes are vaccinated at a social optimum compared to the PNE under true expectation minimizers, and when $c \geq w(c)$.

Following Proposition \ref{proposition:NE_threshold} and Proposition \ref{proposition:th_opt}, we can express the populations of unprotected nodes at the PNE and a social optimum as
\begin{align*}
\xv^*_\UT &:= \{m_1,m_2,\ldots,x_{d_x},0,\ldots,0\}, 
\\ \yv^*_\UT &:= \{m_1,m_2,\ldots,y_{d_y},0,\ldots,0\},
\end{align*}
respectively, where $x_{d_x} \in (0,m_{d_x}]$ and $y_{d_y} \in (0,m_{d_y}]$. We have the following relationship between $\xv^*_\UT$ and $\yv^*_\UT$.

\begin{proposition}\label{proposition:opt_ne}
Let the players be either true expectation minimizers or the vaccination cost and the nonlinear probability weighting function satisfy $c \geq w(c)$. Let social states at the PNE and a social optimum be $\xv^*$ and $\yv^*$. Then, $\yv^* \preceq \xv^*$, or equivalently, either $d_y < d_x$ or $d_y = d_x$ and $y_{d_y} \leq x_{d_x}$. 
\end{proposition}
\begin{IEEEproof}
We prove the result for equilibria under true expectation minimizers. The result for nonlinear probability weighting then follows from Proposition \ref{proposition:threshold_comparison}. 

We drop the superscript in the proof for better readability. Assume on the contrary that $\xv \prec \yv$. Following Lemma \ref{lemma:v_monotone}, we have $v(\yv) > v(\xv)$; recall that $v(\xv)$ is the probability that a randomly chosen neighbor is infected. Let $\Delta \yv := \yv_\UT - \xv_\UT > 0$ (the inequality being component-wise) denote the population of nodes who are unprotected at the social optimum $\yv$, but vaccinated at the PNE. We compute the difference of social costs at the PNE and the social optimum, $\Psi(\xv) - \Psi(\yv)$, as
\begin{align}
& \sum_{d \in \DD} x_{d,\UT} \left[\frac{dv(\xv)}{\delta+dv(\xv)} -c\right]  - \sum_{d \in \DD} y_{d,\UT} \left[\frac{dv(\yv)}{\delta+dv(\yv)} -c\right] \nonumber
\\ = & \sum_{d \in \DD} x_{d,\UT} \left[\frac{dv(\xv)}{\delta+dv(\xv)} -\frac{dv(\yv)}{\delta+dv(\yv)}\right] \nonumber
\\ & \qquad - \sum_{d \in \DD} \Delta y_{d,\UT} \left[\frac{dv(\yv)}{\delta+dv(\yv)} -c\right] \nonumber
\\ < & - \sum_{d \in \DD} \Delta y_{d,\UT} \left[\frac{dv(\yv)}{\delta+dv(\yv)} -c\right]. \label{eq:opt_ne_int}
\end{align} 
It remains to show that the right hand side in \eqref{eq:opt_ne_int} is negative, which will contradict the optimality of $\yv$. We consider the following three exhaustive cases. 
\\ \noindent {\bf Case 1:} $d_y = d_x, y_{d_y} > x_{d_x}$.
In this case, we necessarily have $x_{d_x} < m_{d_x}$. Then, from the definition of the PNE \eqref{eq:PNE_leq} and \eqref{eq:PNE_geq}, we have
$$ c = \frac{d_x v(\xv)}{\delta+d_x v(\xv)} < \frac{d_x v(\yv)}{\delta+d_x v(\yv)}. $$
Furthermore, $\Delta \yv = \{0,\ldots,0,y_{d_y} - x_{d_x},0,\ldots,0\}$ in this case. Therefore, the expression in \eqref{eq:opt_ne_int} is negative.
\\ \noindent {\bf Case 2:} $d_y > d_x,x_{d_x} = m_{d_x}$. 
Here no node with degree strictly larger than $d_x$ remains unprotected at the PNE, i.e., 
$$ c \leq \frac{d' v(\xv)}{\delta+d' v(\xv)} < \frac{d' v(\yv)}{\delta+d'v(\yv)},$$
for every $d' > d_x$. Furthermore, in this case, $\Delta \yv = \{0,\ldots,0,m_{d_x+1},\ldots,y_{d_y},0,\ldots,0\}$. Therefore, the expression in \eqref{eq:opt_ne_int} is negative in this case as well.
\\ \noindent {\bf Case 3:} $d_y > d_x,x_{d_x} < m_{d_x}$. 
Following similar reasoning as the first two cases, we can conclude that
$c < \frac{d' v(\yv)}{\delta+d'v(\yv)}$ for every $d' \geq d_x$. Accordingly, the expression in \eqref{eq:opt_ne_int} is negative. Thus, we have the desired contradiction. 
\end{IEEEproof}

Thus, the above result shows that fewer nodes purchase vaccination under decentralized decision making, i.e., nodes exhibit free riding behavior. Related literature such as \cite{omic2009protecting,trajanovski2015decentralized,la2016interdependent} have also shown that PNEs are inefficient in their respective network security game settings. Our result in Proposition \ref{proposition:opt_ne} is analogous to those results. In addition, Proposition \ref{proposition:opt_ne} holds even under behavioral probability weighting when the vaccination cost is sufficiently high. While it is challenging to compare the social costs under decentralized and centralized decision making in general networks and probability weighting functions, we prove an upper bound on the difference of social costs for true expectation minimizers.

\begin{figure*}[t]
    \begin{subfigure}[t]{0.32\textwidth}
        \centering
        \includegraphics[width=\linewidth]{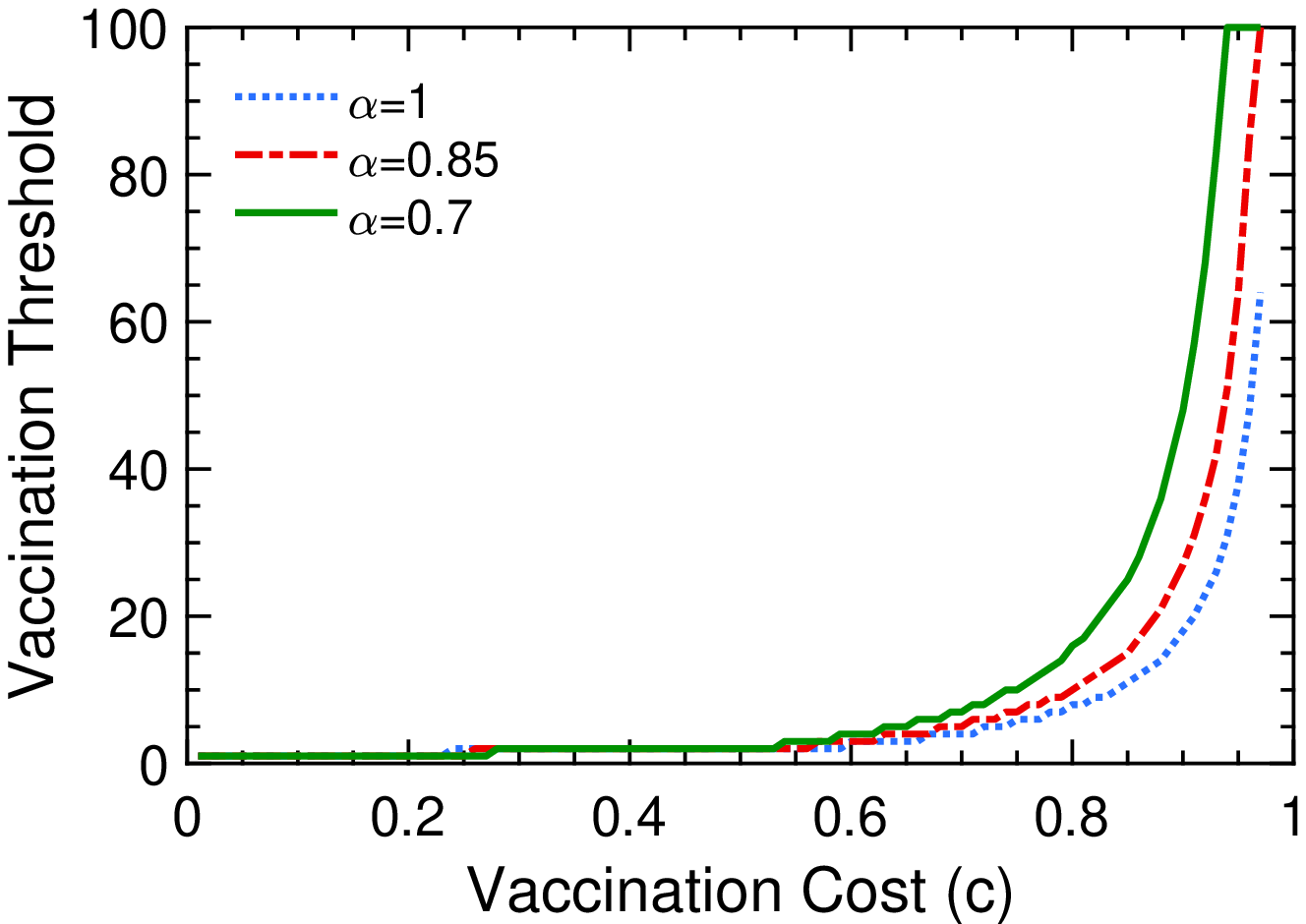}
        \caption{Comparison of vaccination thresholds at the pure Nash equilibrium.}
        \label{fig:th_power}
    \end{subfigure}
    \hspace*{\fill}
    \begin{subfigure}[t]{0.32\textwidth}
        \centering
        \includegraphics[width=\linewidth]{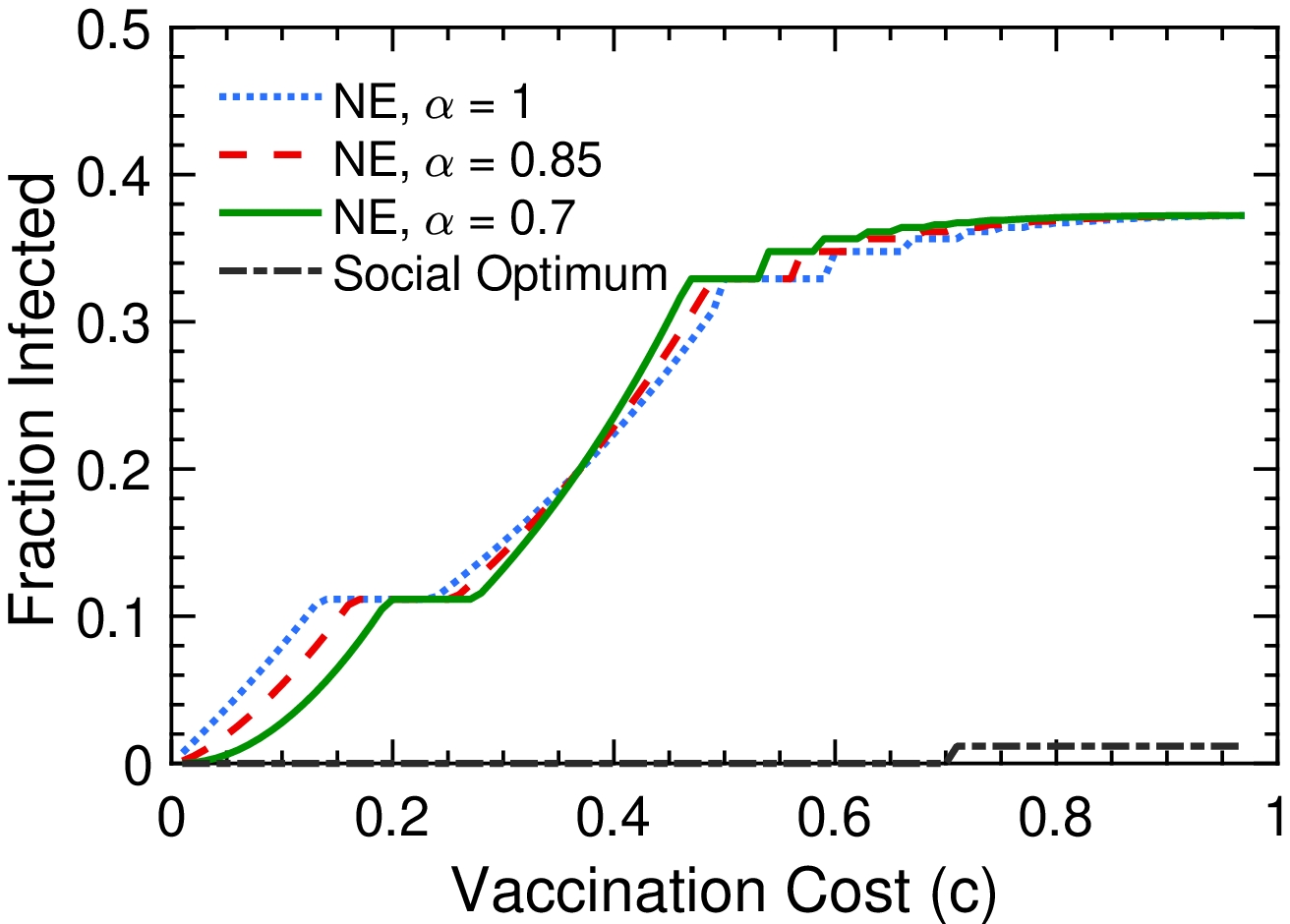}
        \caption{Comparison of expected fractions of infected nodes.}
        \label{fig:frac_power}
    \end{subfigure}
    \hspace*{\fill}
    \begin{subfigure}[t]{0.32\textwidth}
        \centering
        \includegraphics[width=\linewidth]{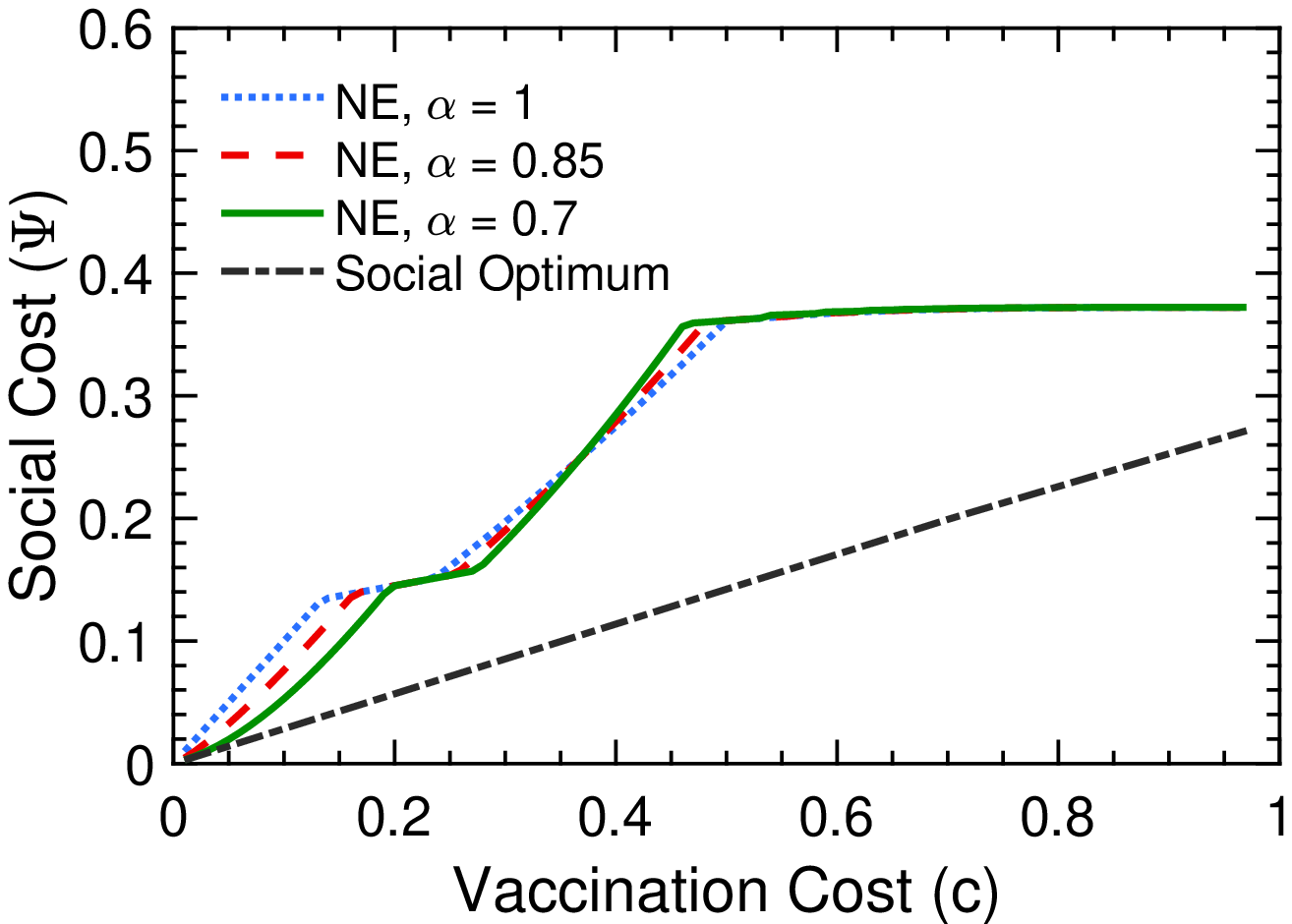}
        \caption{Comparison of social costs.}
        \label{fig:sc_power}
    \end{subfigure}
    \caption{PNE thresholds, expected fractions of infected nodes, and social costs for true and behavioral perceptions
of infection probabilities (with Prelec weighting functions) in networks with a power-law degree distribution with exponent $3$.}
\end{figure*}

\begin{proposition}\label{proposition:opt_diff_bound}
Let the players be true expectation minimizers. Let social states at the PNE and a social optimum be $\xv^*$ and $\yv^*$, respectively. Then, $\Psi(\xv^*) - \Psi(\yv^*) \leq \frac{\langle d\rangle}{\delta}$.
\end{proposition}
\begin{IEEEproof}
We drop the superscript $*$ from the social states for better readability. From Proposition \ref{proposition:opt_ne}, we know that $\yv \preceq \xv$. Then, we have $v(\yv) \leq v(\xv)$ following Lemma \ref{lemma:v_monotone}. Let $\Delta \xv := \xv_{\UT} - \yv_{\UT} \geq 0$ (the inequality being component wise) denote the population of nodes who are vaccinated at the social optimum, but not at the PNE. We now compute
\begin{align}
& \Psi(\xv) - \Psi(\yv) = \sum_{d \in \DD} (y_{d,\UT} + \Delta x_{d,\UT}) \left[\frac{dv(\xv)}{\delta+dv(\xv)} -c\right] \nonumber
\\ & \qquad \qquad \qquad - \sum_{d \in \DD} y_{d,\UT} \left[\frac{dv(\yv)}{\delta+dv(\yv)} -c\right] \nonumber
\\ & = \sum_{d \in \DD} \frac{y_{d,\UT}dv(\xv)}{\delta+dv(\xv)} -\frac{y_{d,\UT}dv(\yv)}{\delta+dv(\yv)} + \Delta x_{d,\UT} \left[\frac{dv(\xv)}{\delta+dv(\xv)} -c\right] \nonumber
\\ & \leq \sum_{d \in \DD} y_{d,\UT} \left[\frac{dv(\xv)}{\delta+dv(\xv)} -\frac{dv(\yv)}{\delta+dv(\yv)}\right] \label{eq:int_diff_1}
\\ & \leq \sum_{d \in \DD} \frac{\delta y_{d,\UT} d (v(\xv)-v(\yv))}{(\delta+dv(\yv))^2} \leq \sum_{d \in \DD} \frac{m_d d}{\delta} = \frac{\langle d\rangle}{\delta}, \label{eq:int_diff_3}
\end{align}
where \eqref{eq:int_diff_1} follows from the definition of the PNE \eqref{eq:PNE_leq},\footnote{Note that $\Delta x_{d,\UT} > 0$ only if a nonzero fraction of degree $d$ nodes remain unprotected. For such nodes, the infection probability at the endemic state must be at most the vaccination cost $c$.} and the inequalities in \eqref{eq:int_diff_3} hold since $v(\yv) \leq v(\xv)$, $v(\xv) \leq 1, v(\yv) \geq 0$, and $y_{d,\UT} \leq m_d, \forall d \in \DD$. 
\end{IEEEproof}

While the above bound is not necessarily tight, it provides an important insight: the difference of social costs under decentralized and centralized decision making does not grow faster than the mean degree of the network.

\section{Numerical Illustrations}

We now illustrate some of our theoretical findings via numerical examples. We consider networks with a power-law degree distribution with exponent $3$ and $\DD = \{1,2,\ldots,100\}$. We choose the curing rate $\delta=2$. We compute the PNE and social optimum social states by leveraging threshold properties and the characteristics of candidate social states.  All computations were carried out in MATLAB.

We first compare the PNE thresholds for different values of the Prelec weighting parameter $\alpha$ (equation \eqref{eq:prelec}). Recall that $\alpha=1$ corresponds to true expectation minimizers and smaller values of $\alpha$ implies a higher degree of over and underweighting of low and high probabilities, respectively. Figure \ref{fig:th_power} shows that for vaccination costs up to $0.5$, the PNE thresholds under true and behavioral perceptions of probabilities are almost identical. As $c$ increases to $1$, the threshold is much higher under nonlinear probability weighting than under true expectation minimizers, confirming our findings in Propositions \ref{proposition:threshold_comparison} and \ref{proposition:powerlaw3}. 

Although the PNE thresholds are essentially identical for vaccination costs up to $0.5$, the population of unprotected nodes are quite different for different values of $\alpha$. As shown in Proposition \ref{proposition:threshold_comparison}, for $c < \frac{1}{e}$, a larger fraction of nodes are vaccinated under behavioral probability weighting than true expectation minimizers, and vice versa. This can be observed in Figure \ref{fig:frac_power} which shows that the expected fraction of infected nodes at the PNE (i.e., the quantity $\sum_{d \in \DD} x_{d,\UT} \frac{dv(\xv)}{\delta+dv(\xv)}$) is smaller for smaller values of $\alpha$ when $c < \frac{1}{e}$, and vice versa. Furthermore, as $c \to 1$, the expected fraction of infected nodes saturates for the PNEs. This is because the total fraction of nodes with degrees larger than $10$ is negligible ($0.003$), and increase in vaccination threshold beyond this level has limited impact on the spread of the epidemic. 

On the other hand, the expected fraction of infected nodes at the social optimum is negligible and exhibits only slight increase as the vaccination cost increases. We observed that at the social optimum, the vaccination threshold is $1$ (with a non-negligible fraction of nodes with degree $1$ being vaccinated) for the entire range of $c$, in contrast with the PNE. Finally, Figure \ref{fig:sc_power} compares the respective social costs. Note that the social cost at the social optimum is dominated by the vaccination cost, while at the equilibrium, $\Psi$ is dominated by the expected fraction of nodes that are infected.

\section{Conclusion}
We investigated decentralized vaccination decisions by human decision-makers against networked SIS epidemics in a population game framework. We first established the existence and uniqueness of a threshold equilibrium, where only nodes with degrees larger than a threshold vaccinate, and nodes with degrees smaller than the threshold remain unprotected. We then showed that if the vaccination cost is larger than a quantity that depends on the probability weighting function, then behavioral biases cause fewer nodes to vaccinate, and vice versa. Furthermore, this result holds irrespective of the network topology. We then obtained tight bounds on the ratio of vaccination thresholds under nonlinear and true perceptions of probabilities for a class of networks whose degree distribution follows a power-law. Finally, we analyzed the socially optimal vaccination policy. When nodes perceive probabilities as their true values, we showed that fewer nodes vaccinate at a PNE compared to a socially optimal vaccination policy, and the difference in social costs is upper bounded by a quantity proportional to the mean degree of the network. Our numerical illustrations provided additional insights into the vaccination decisions and security levels of networks under decentralized and human decision-making.

\bibliographystyle{IEEEtran}
\bibliography{refs_new,refs}

\appendix
\subsection{Proof of Theorem \ref{theorem:app_bullo}}\label{appendix_proof}

According to the NIMFA approximation of the SIS epidemic \cite{van2009virus}, the infection probability of node $i$, $z_i(t) \in [0,1], t \geq 0$, evolves as
\begin{align}\label{eq:app_nimfa}
\frac{d z_i(t)}{dt} & = -\delta z_i(t) + (1-z_i(t)) \sum_{j \in \mathcal{N}_i} a_{ji}z_j(t),
\end{align}
where $\mathcal{N}_i$ is the set of in-neighbors of node $i$ and $a_{ji} \in \Rb$ denotes the weight or strength of connection between nodes $j$ and $i$. The above dynamics can be written in vector form as
\begin{align}\label{eq:app_nimfa_linear}
\frac{d z(t)}{dt} & = -\Delta z(t)+(\mathbb{I}-\mathtt{diag}(z(t)) Az(t),
\end{align}
where $\mathbb{I}$ is the identity matrix, $\Delta = \mathtt{diag}(\delta,\delta,\ldots,\delta)$ is the diagonal matrix of all curing rates, and $A$ is the weighted adjacency matrix of the network. In \cite{bullo2016lectures,khanafer2016stability}, it was shown that the statement of Theorem \ref{theorem:app_bullo} holds for the NIMFA dynamics \eqref{eq:app_nimfa} for $R$ being the spectral radius of the matrix $\Delta^{-1}A^{T}$. We now prove Theorem \ref{theorem:app_bullo} using the result from \cite{bullo2016lectures,khanafer2016stability} by establishing a connection between the NIMFA and DBMF approximations (\eqref{eq:app_nimfa} and \eqref{eq:app_dbmf}, respectively).

\noindent {\bf Proof of Theorem \ref{theorem:app_bullo}:} If all nodes vaccinate, then $R(\xv)=0 \leq 1$, and clearly the only endemic state is $p_d(\xv)=0, \forall d \in \DD$. Now, consider a social state $\xv$ such that at least some fraction of nodes remain unprotected. We now construct a directed graph $\hat{\GG}(\xv)$ from the original network $\GG$ such that the NIMFA of the SIS epidemic on $\hat{\GG}(\xv)$ coincides with the DBMF approximation (stated in \eqref{eq:app_dbmf}) on $\GG$. Specifically, let $\hat{\GG}(\xv)$ be a weighted directed graph with the set of nodes $\DD$; each node in $\hat{\GG}(\xv)$ represents a degree present in $\GG$. 

We define the weight of an edge $(j,i) \in \DD^2$ as $a_{ij}(\xv) := i\hat{q}_j(\xv)$, where $\hat{q}_j(\xv) = \frac{jx_{j,\UT}}{\langle d\rangle}$. In particular, the strength of the directed edge $(j,i)$ is the product of $i$ and the probability that a node with degree $i$ encounters an unprotected node with degree $j$ among her neighbors in the original graph $\GG$. Accordingly, the adjacency matrix of $\hat{\GG}(\xv)$ is given by $\hat{A}(\xv) = \mathbf{d}\cdot\mathbf{q}(\xv)^T$, where $\mathbf{d}$ is the vector of degrees of $\GG$, and $\mathbf{q}(\xv) = [\frac{jx_{j,\UT}}{\langle d\rangle}]_{j \in \mathcal{D}}$. In other words, $\hat{A}(\xv)$ has rank one with the only nonzero eigenvalue given by $\hat{\lambda}(\xv) = \mathbf{q}(\xv)^T\mathbf{d} = \sum_{i\in\mathcal{D}} i\hat{q}_i(\xv)$. 

Note that the evolution of the SIS epidemic in $\hat{\GG}(\xv)$ under the NIMFA \eqref{eq:app_nimfa} coincides with \eqref{eq:app_dbmf}. Furthermore, the spectral radius of the matrix $\Delta^{-1}\hat{A}^T(\xv)$ is $R(\xv) =\sum_{i\in\mathcal{D}} \frac{i\hat{q}_i(\xv)}{\delta}$. Therefore, the uniqueness and stability of the equilibrium points of the dynamics in \eqref{eq:app_dbmf} follow directly from \citep{bullo2016lectures,khanafer2016stability}. \hfill \IEEEQED

\subsection{Proof of Lemma \ref{lemma:th_bound}}
\label{appendix:lemma}

\noindent {\bf Proof of Lemma \ref{lemma:th_bound}:} Recall that $\beta \in [2,3]$. Therefore, $\frac{1}{i^{\beta-2}(\delta+iv_{\beta}(\xv_t))}$ in \eqref{eq:th_v} is monotonically decreasing in $i$, and
\begin{equation}\label{eq:It_bounds}
\begin{aligned}
& \sum^t_{i=d_0} \frac{1}{i^{\beta-2}(\delta+iv_{\beta}(\xv_t))} 
\\ & \qquad \geq \sum^t_{i=d_0} \frac{1}{i(\delta+iv_{\beta}(\xv_t))} \geq \int_{d_0}^{t} \frac{di}{i(\delta+iv_{\beta}(\xv_t))},
\end{aligned}
\end{equation}
where $i$ corresponds to the degree and $d$ refers to the integration variable. We denote the R.H.S. by
\begin{align*}
\mathcal{I}(t) & := \int_{d_0}^{t} \frac{di}{i(\delta+iv_{\beta}(\xv_t))} = \frac{1}{\delta} \int_{d_0}^{t} \frac{(\delta + iv_{\beta}(\xv_t) - iv_{\beta}(\xv_t)) di}{i(\delta+iv_{\beta}(\xv_t))} 
\\ & = \frac{1}{\delta} \left[ \int_{d_0}^{t} \frac{di}{i} - \int_{d_0}^{t} \frac{v_{\beta}(\xv_t) di}{\delta+iv_{\beta}(\xv_t)} \right] 
\\ & = \frac{1}{\delta} \left[ \log\left(\frac{t}{d_0}\right) - \log\left(\frac{\delta + tv_{\beta}(\xv_t)}{\delta + d_0v_{\beta}(\xv_t)}\right) \right] 
\\ & = \frac{1}{\delta} \log\left( \frac{t(\delta+d_0v_{\beta}(\xv_t))}{d_0(\delta+tv_{\beta}(\xv_t))} \right).
\end{align*}
Recall from \eqref{eq:th_v} that the first term in \eqref{eq:It_bounds} is $\frac{\langle d \rangle}{\kappa}$. Therefore, from $\mathcal{I}(t) \leq \frac{\langle d \rangle}{\kappa}$, we obtain
\begin{align}
& \log\left( \frac{t(\delta+d_0v_{\beta}(\xv_t))}{d_0(\delta+tv_{\beta}(\xv_t))} \right) \leq \frac{\delta \langle d \rangle}{\kappa} \nonumber
\\ \implies & \frac{t(\delta+d_0v_{\beta}(\xv_t))}{d_0(\delta+tv_{\beta}(\xv_t))} - 1 \leq e^{\frac{\delta \langle d \rangle}{\kappa}} - 1 = B_1 - 1 \nonumber 
\\ \implies & \frac{(B_1-1)^{-1}(t-d_0)\delta}{d_0} \leq \delta+tv_{\beta}(\xv_t) \label{eq:tvt_upper}
\\ \implies & \frac{\delta}{\delta+t v_{\beta}(\xv_t)} \leq \frac{d_0 (B_1-1)}{t-d_0}
\\ \implies & \frac{tv_{\beta}(\xv_t)}{\delta+tv_{\beta}(\xv_t)} \geq 1 - \frac{d_0(B_1-1)}{t-d_0} = \frac{t-d_0B_1}{t-d_0}. \label{eq:th_lower1}
\end{align}

For the second part, let $d_0 > 1$ and $\beta=3$. Then, 
\begin{equation*}
\begin{aligned}
\frac{\langle d \rangle}{\kappa} = & \sum^t_{i=d_0} \frac{1}{i^{\beta-2}(\delta+iv_{\beta}(\xv_t))} \leq \int_{d_0-1}^{t} \frac{di}{i(\delta+iv_{\beta}(\xv_t))}.
\end{aligned}
\end{equation*}
Thus, we repeat the above analysis reversing the inequalities and replacing $d_0$ with $d_0-1$, and obtain
\begin{equation}\label{eq:part2_lower1}
\frac{tv_{3}(\xv_t)}{\delta+tv_{3}(\xv_t)} \leq \frac{t-(d_0-1)B_1}{t-(d_0-1)}. 
\end{equation}
This concludes the proof.  \hfill \IEEEQED
\end{document}